\documentclass[twocolumn,aps,prc,superscriptaddress,showpacs,floatfix]{revtex4}
\usepackage{graphicx}

\begin{document}

\title{System size dependence of elliptic flows in relativistic heavy-ion
collisions}
\author{Lie-Wen Chen}
\affiliation{Institute of Theoretical Physics, Shanghai Jiao Tong University, Shanghai
200240, China}
\affiliation{Center of Theoretical Nuclear Physics, National Laboratory of Heavy Ion
Accelerator, Lanzhou 730000, China}
\author{Che Ming Ko}
\affiliation{Cyclotron Institute and Physics Department, Texas A\&M University, College
Station, Texas 77843-3366}
\date{\today }

\begin{abstract}
The elliptic flows in both Cu+Cu and Au+Au collisions at the Relativistic
Heavy Ion Collider are studied in a multi-phase transport model. For both
collisions at same reduced impact parameter and minimum bias collisions, the
elliptic flow of partons in Cu+Cu collisions is about a factor of three
smaller than that in Au+Au collisions at same energy. The reduction factor
is similar to the ratio of the sizes of the two colliding systems and is
also related to the combined effects of initial energy density and spatial
elliptic deformation in the two reactions. A similar system size dependence
is also seen in the elliptic flow of charged hadrons from minimum bias
collisions.
\end{abstract}

\maketitle

\section{Introduction}

There have been extensive studies on the azimuthal anisotropy of hadron
momentum distributions in the transverse plane perpendicular to the beam
direction, particularly the elliptic flow $v_{2}$, in heavy-ion collisions
at various energies \cite{reisdorf}. The hadron transverse momentum
anisotropy is generated by the pressure anisotropy in the initial compressed
matter formed in non-central heavy ion collisions \cite{Barrette94, Appel98}
and thus depends on the initial geometry and energy density of a collision
as well as the properties of produced matter during the early stage of the
collision. For heavy-ion collisions at the Relativistic Heavy Ion Collider (%
\textrm{RHIC}), it has been shown that not only the larger elliptic flow 
\cite{Ollit92,Rqmd,Danie98,Zhang99,Zheng99,Voloshin03} but also the smaller
higher-order anisotropic flows \cite%
{Kolb99,Teaney99,Kolb00,Kolb03,STAR03,chen04,Kolb04} are sensitive to the
properties of initial dense matter. To study the dependence of the hadron
anisotropic flow on the initial geometry and energy density in a heavy ion
collision, one can vary the impact parameter of the collision or the atomic
number of colliding nuclei. Although there were already extensive studies on
the dependence of elliptic flow on the impact parameter of heavy ion
collisions, its dependence on the size of colliding nuclei has only been
started recently \cite{manly}.

In the present work, we use a multi-phase transport (\textrm{AMPT}) model,
that includes both initial partonic and final hadronic interactions \cite%
{Zhang:2000bd,Lin:2001cx}, to study the elliptic flow in both Cu+Cu and
Au+Au collisions at $\sqrt{s}=200$ \textrm{AGeV} at \textrm{RHIC}. In
particular, we use the version with string melting, i.e., allowing hadrons
that are expected to be formed from initial strings to convert to their
valence quarks and antiquarks \cite{Lin:2001zk,LinHBT02,ko}, which has been
shown to be able to explain the measured $p_{T}$ dependence of $v_{2}$ and $%
v_{4}$ of mid-rapidity charged hadrons with a parton scattering cross
section of about $10$ \textrm{mb}. We find that the elliptic flow scales
approximately with the size of the colliding system and is also proportional
to the product of the spatial elliptic deformation and energy density during
the initial stage of a collision.

This paper is organized as follows. In Sec. \ref{ampt}, the \textrm{AMPT}
model is briefly reviewed. Results for the parton elliptic flow and spatial
elliptic deformation as well as their system size dependence are shown in
Sec. \ref{v2s2}, while the system size dependence of the elliptic flow of
charged hadrons is shown in Sec. \ref{size}. Finally, a brief summary is
given in Sec. \ref{summary}.

\section{The AMPT model}

\label{ampt}

The \textrm{AMPT} model \cite{Zhang:2000bd,Lin:2001cx,zhang,pal,ampt} is a
hybrid model that uses minijet partons from hard processes and strings from
soft processes in the Heavy Ion Jet Interaction Generator (\textrm{HIJING})
model \cite{Wang:1991ht} as the initial conditions for modeling heavy ion
collisions at ultra-relativistic energies. Since the initial energy density
in Au+Au collisions at \textrm{RHIC} is much larger than the critical energy
density at which the hadronic matter to quark-gluon plasma transition would
occur \cite{Kharzeev:2001ph,zhang}, we use the version which allows the
melting of initial excited strings into partons \cite{Lin:2001zk}. In this
string melting scenario, hadrons (mostly pions), that would have been
produced from string fragmentation, are converted instead to valence quarks
and/or antiquarks with current quark masses. Interactions among these
partons are described by Zhang's\ parton cascade (\textrm{ZPC}) model \cite%
{Zhang:1997ej}. At present, this model includes only parton-parton elastic
scatterings with an in-medium cross section given by: 
\begin{equation}
\frac{d\sigma _{p}}{dt}\approx \frac{9\pi \alpha _{s}^{2}}{2}\left( 1+{\frac{%
{\mu ^{2}}}{s}}\right) \frac{1}{(t-\mu ^{2})^{2}},  \label{crscp}
\end{equation}%
where the strong coupling constant $\alpha _{s}$ is taken to be $0.47$, and $%
s$ and $t$ are usual Mandelstam variables. The effective screening mass $\mu 
$ depends on the temperature and density of the partonic matter but is taken
as a parameter in \textrm{ZPC} for fixing the magnitude and angular
distribution of parton scattering cross section. Since there are no
inelastic scatterings, only quarks and antiquarks from the melted strings
are present in the partonic matter. The transition from the partonic matter
to the hadronic matter is achieved using a simple coalescence model, which
combines two nearest quark and antiquark into mesons and three nearest
quarks or antiquarks into baryons or anti-baryons that are close to the
invariant mass of these partons. The present coalescence model is thus
somewhat different from the ones recently used extensively \cite%
{greco,hwa,fries,molnar03} for studying hadron production at intermediate
transverse momenta. The final-state hadronic scatterings are then modeled by
a relativistic transport (\textrm{ART}) model \cite{Li:1995pr}. Using parton
scattering cross sections of $6$-$10$ \textrm{mb}, the \textrm{AMPT} model
with string melting is able to reproduce both the centrality and transverse
momentum (below $2$ \textrm{GeV}$/c$) dependence of the elliptic flow \cite%
{Lin:2001zk} and pion interferometry \cite{LinHBT02} measured in Au+Au
collisions at $\sqrt{s}=130$ \textrm{AGeV} at \textrm{RHIC} \cite%
{Ackermann:2000tr,STARhbt01}. It has also been used for studying the kaon
interferometry \cite{lin} and the elliptic flow of charmed mesons \cite%
{charm} in these collisions. We note that the above cross sections are
significantly smaller than that needed to reproduce the parton elliptic flow
from the hydrodynamic model \cite{molnar}. The resulting hadron elliptic
flows in the \textrm{AMPT} model with string melting are, however, amplified
by modeling hadronization via quark coalescence \cite{molnar03}, leading to
a satisfactory reproduction of experimental data.

\section{Elliptic flow and spatial anisotropy of partons}

\label{v2s2}

The anisotropic flows $v_{n}$ of particles are the Fourier coefficients in
the decomposition of their transverse momentum spectra in the azimuthal
angle $\phi $ with respect to the reaction plane \cite{Posk98}, i.e., 
\begin{equation}
E\frac{d^{3}N}{dp^{3}}=\frac{1}{2\pi }\frac{dN}{p_{T}dp_{T}dy}%
[1+\sum_{n=1}^{\infty }2v_{n}(p_{T},y)\cos (n\phi )]
\end{equation}%
Because of the symmetry $\phi \leftrightarrow -\phi $ in collision geometry,
no sine terms appear in above expansion. For particles at midrapidity in
collisions with equal mass nuclei, anisotropic flows of odd orders vanish as
a result of the additional symmetry $\phi \leftrightarrow \phi +\pi $.
Anisotropic flows generally depend on particle transverse momentum and
rapidity, and for a given rapidity the anisotropic flows at transverse
momentum $p_{T}$ can be evaluated according to 
\begin{equation}
v_{n}(p_{T})=\left\langle \cos (n\phi )\right\rangle ,
\end{equation}%
where $\left\langle \cdot \cdot \cdot \right\rangle $ denotes average over
the azimuthal distribution of particles with transverse momentum $p_{T}$.
The elliptic flow $v_{2}$ can further be expressed in terms of
single-particle averages: 
\begin{equation}
v_{2}(p_{T})=\left\langle \frac{p_{x}^{2}-p_{y}^{2}}{p_{x}^{2}+p_{y}^{2}}%
\right\rangle
\end{equation}%
where $p_{x}$ and $p_{y}$ are, respectively, projections of the particle
momentum in and perpendicular to the reaction plane.

Since the \textrm{AMPT} model also provides information on the spatial
anisotropy of colliding matter, which is responsible for generating the
momentum anisotropic flows, it is of interest to introduce the spatial
anisotropic coefficient $s_{n}$ by expressions similar to those for the
anisotropic flows $v_{n}$ but in terms of the spatial distributions of
particles in the transverse plane. In particular, the spatial elliptic
deformation $s_{2}$ or eccentricity can be obtained from%
\begin{equation}
s_{2}=\left\langle \frac{x^{2}-y^{2}}{x^{2}+y^{2}}\right\rangle,
\end{equation}%
where $x$ and $y$ are, respectively, projections of the particle coordinate
in and perpendicular to the reaction plane.

\begin{figure}[th]
\includegraphics[scale=1.05]{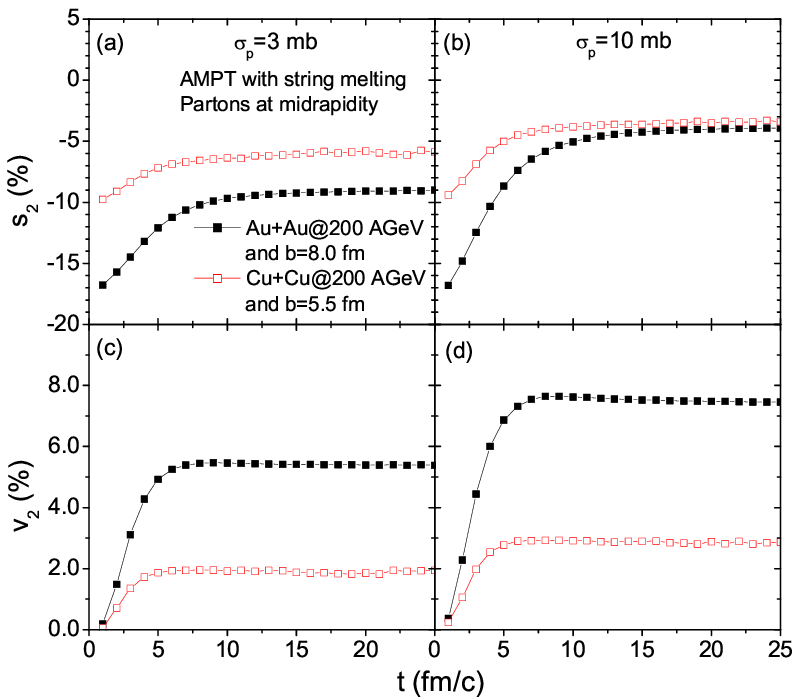}
\caption{{\protect\small (Color online) Time evolution of spatial elliptic
deformation }$s_{2}${\protect\small \ and momentum elliptic flow }$v_{2}$%
{\protect\small \ of partons at midrapidity in Cu+Cu collisions at }$b=5.5$%
{\protect\small \ fm (open squares) and Au + Au collisions at }$b=8$%
{\protect\small \ fm (solid squares) with the same energy per nucleon }$%
\protect\sqrt{s}=200${\protect\small \ AGeV for parton scattering cross
sections }$\protect\sigma _{p}=3${\protect\small \ mb (left panels) and }$%
\protect\sigma _{p}=10${\protect\small \ mb (right panels).}}
\label{vs2time}
\end{figure}

Using the \textrm{AMPT} model in the string melting scenario with parton
scattering cross sections $\sigma _{p}=3$ and $10$ \textrm{mb}, we first
study the time evolution of $s_{2}$ and $v_{2}$ of partons at midrapidity in
Cu+Cu collisions at $b=5.5$ \textrm{fm} and Au+Au collisions at $b=8$ 
\textrm{fm} with the same energy per nucleon, i.e., $\sqrt{s}=200$ \textrm{%
AGeV}, and the results are shown in Fig. \ref{vs2time}. The two impact
parameters, which are similar to the average values in minimum bias
collisions, correspond to the same reduced impact parameter $b/b_{\mathrm{max%
}}$, where $b_{\mathrm{max}}$ is the sum of the radii of colliding nuclei.
They would lead to same $s_{2}$ if the two nuclei were sharp spheres.
Because of diffused surfaces, a much large $s_{2}$ is, however, produced in
Au+Au collisions than in Cu+Cu collisions as shown in Figs. \ref{vs2time}(a)
and  \ref{vs2time}(b). For both collisions, the spatial elliptic deformation 
$s_{2}$ is initially large and decreases with time. It reaches the
saturation value at times which depend on both the size of the reaction
system and the parton scattering cross section. Specifically, the $s_{2}$ in
Cu+Cu collisions has an earlier saturation time, implying that the fireball
(or quark gluon plasma) formed in a smaller reaction system has a shorter
lifetime. This behavior is understandable since the larger reaction system
has a larger initial parton number density in the transverse plane and thus
takes longer time to freeze out. The nonzero spatial elliptic deformation $%
s_{2}$ indicates that the parton spatial distribution is nonspherical at
freeze out, with the larger parton scattering cross section leading to a
smaller spatial anisotropy.

Time evolution of the elliptic flow $v_{2}$ of partons is shown in Figs. \ref%
{vs2time}(c) and \ref{vs2time}(d) for the two parton scattering cross
sections of $3$ and $10$ \textrm{mb}, respectively. The elliptic flow $v_{2}$
is seen to saturate earlier in the collisions, and the saturation time
depends on the size of the reaction system, i.e., at about $5$ \textrm{fm/}$c
$ for Cu+Cu collisions while about $7$ \textrm{fm/}$c$ for Au+Au collisions.
This is similar for both parton scattering cross sections, except that the
larger one leads to a larger elliptic flow. The shorter saturation time of $%
v_{2}$ in lighter reaction system is consistent with the earlier saturation
time of $s_{2}$ seen in Figs. \ref{vs2time}(a) and \ref{vs2time}(b) for that
system. An interesting result predicted by the \textrm{AMPT} model is that
the parton $v_{2}$ in Cu+Cu collisions is significantly smaller than that in
Au+Au collisions. For instance, the final values of $v_{2}$ of partons are
about $1.9\%$ and $2.9\%$ , respectively, for $\sigma _{p}=3$ \textrm{mb}
and $10$ \textrm{mb} in Cu+Cu collisions, and they increase to about $5.4\%$
and $7.5\%$ , respectively, for $\sigma _{p}=3$ \textrm{mb} and $10$ \textrm{%
mb} in Au+Au collisions. For both cross sections, the strength of parton $%
v_{2}$ is enhanced by a factor of about 3 from Cu+Cu to Au+Au collisions at
same reduced impact parameter, and is roughly similar to the ratio of the
sizes of Au and Cu nuclei.

\begin{figure}[th]
\includegraphics[scale=1.15]{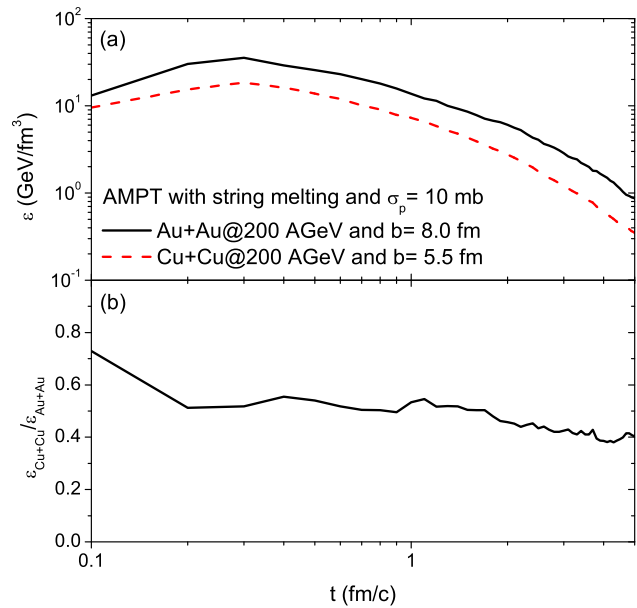}
\caption{{\protect\small (Color online) Time evolution of energy densities
of midrapidity partons (upper panel) and their ratio (lower panel) in Cu+Cu
collisions at }$b=5.5${\protect\small \ fm (dashed line) and Au + Au
collisions at }$b=8${\protect\small \ fm (solid line) with the same energy
per nucleon }$\protect\sqrt{s}=200${\protect\small \ AGeV for parton
scattering cross sections }$\protect\sigma _{p}=10${\protect\small \ mb.}}
\label{energy}
\end{figure}

The magnitude of elliptic flow is affected not only by the spatial elliptic
deformation but also by the energy density during the initial stage of
collisions. In Fig. \ref{energy}, we show the time evolution of the energy
densities of midrapidity partons (upper panel) and their ratio (lower panel)
in Cu+Cu collisions at $b=5.5$ fm (dashed line) and Au+Au collisions at $b=8$
fm (solid line) for parton scattering cross section of 10 mb. As in Ref. 
\cite{zhang}, the energy density is calculated for partons in the central
cell that has a transverse radius of $1$ fm and a longitudinal dimension of $%
5\%$ of the time $t$ with the time coordinate starting when the two nuclei
fully overlap in the longitudinal direction. It is seen that the initial
energy density in Cu+Cu collisions is about $2/3$ of that in Au+Au
collisions. Combining this effect with that due to initial spatial elliptic
deformation, which has a ratio of about $1/2$ for the two colliding systems,
gives a reduction factor of about $1/3$, which is comparable to the ratio of
the sizes of the two colliding systems. The dependence on the system size
seen in the parton elliptic flows from the \textrm{AMPT} model is thus
related to the combined effects of initial energy density and spatial
elliptic deformation.

\begin{figure}[th]
\includegraphics[scale=0.9]{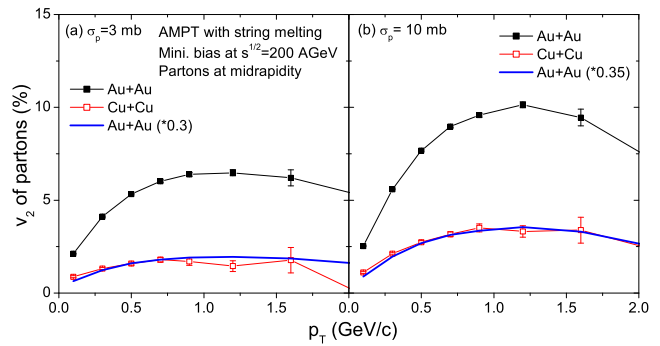}
\caption{{\protect\small (Color online) Transverse momentum }$p_{T}$%
{\protect\small \ dependence of }$v_{2}${\protect\small \ of midrapidity
partons from minimum bias events in Au+Au (solid squares) and Cu+Cu (open
squares) collisions at }$\protect\sqrt{s}=200${\protect\small \ AGeV for
parton scattering cross sections of }$\protect\sigma _{p}=3${\protect\small %
\ mb (left panel) and }$\protect\sigma _{p}=10${\protect\small \ mb (right
panel). The solid line is }$0.3${\protect\small \ (}$0.35${\protect\small )
times }$v_{2}${\protect\small \ of Au+Au for }$\protect\sigma _{p}=3$%
{\protect\small \ (}$10${\protect\small )\ mb.}}
\label{v2PTparton}
\end{figure}

The same system size dependence is also seen in the parton differential
elliptic flow as a function of parton transverse momentum $p_{T}$ for
minimum bias collisions. This is shown in Figs. \ref{v2PTparton}(a) and \ref%
{v2PTparton}(b) for midrapidity partons from minimum bias events in Au+Au
and Cu+Cu collisions at $\sqrt{s}=200$ \textrm{AGeV} for parton scattering
cross sections of $3$ \textrm{mb }and $10$ \textrm{mb}, respectively. As
shown by the solid lines in Figs. \ref{v2PTparton}(a) and \ref{v2PTparton}%
(b), the parton differential $v_{2}$ in Cu+Cu collisions is about $0.3$ ($%
0.35$) times that in Au+Au collisions for parton scattering cross section of 
$3$ ($10$) \textrm{mb}.

We note that a detailed comparison between the parton elliptic flows in the
two colliding systems further indicates that the parton $v_{2}$ in the
lighter reaction system is more sensitive to the parton cross sections. For
example, the parton $v_{2}$ is enhanced by a factor of about $1.53$ in Cu+Cu
collisions but is enhanced by a factor of about $1.39$ in Au+Au collisions,
when the parton cross section is increased by a factor of about $3$.

\section{System size dependence of the charged hadron elliptic flow}

\label{size}

\begin{figure}[th]
\includegraphics[scale=0.9]{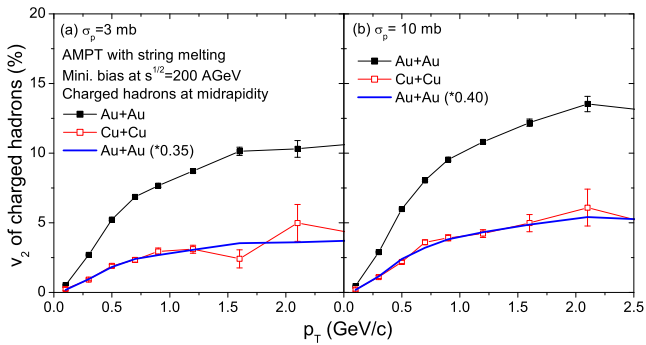}
\caption{{\protect\small (Color online) Same as Fig. \protect\ref{v2PTparton}
but for charged hadrons. The solid line is }$0.35${\protect\small \ (}$0.4$%
{\protect\small ) times }$v_{2}${\protect\small \ of Au+Au for }$\protect%
\sigma _{p}=3${\protect\small \ (}$10${\protect\small )\ mb.}}
\label{v2PTchg}
\end{figure}

The anisotropic flows of partons are transferred to those of hadrons when
the latter are formed from the coalescence of quarks and/or antiquarks.
Although scatterings among hadrons are included in the \textrm{AMPT} model,
they do not affect much the hadron anisotropy flows as a result of the small
spatial anisotropy and pressure in the hadronic matter \cite{Lin:2001zk}. In
Fig. \ref{v2PTchg}, we show the final $v_{2}$ of charged hadrons in the
pseudorapidity range $\left\vert \eta \right\vert <1.2$ in minimum bias
Au+Au and Cu+Cu collisions at $\sqrt{s}=200$ \textrm{AGeV} as functions of
the transverse momentum $p_{T}$ for parton scattering cross sections $\sigma
_{p}=3$ and $10$ \textrm{mb}. It is seen that the charged hadron $v_{2}$ in
the lighter reaction system Cu+Cu is significantly smaller than that in the
heavier reaction system Au+Au. As shown by the solid lines in Figs. \ref%
{v2PTchg}(a) and \ref{v2PTchg}(b), the charged hadron $v_{2}$ in Cu+Cu
collisions is about $0.35$ ($0.4$) times that in Au+Au collisions for parton
scattering cross section of $3$ ($10$) \textrm{mb}. The elliptic flow of
charged hadrons thus shows a similar dependence on the size of the colliding
system as that of partons.

\begin{figure}[th]
\includegraphics[scale=0.9]{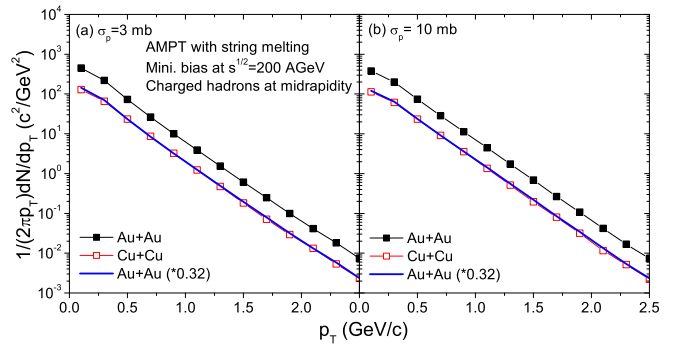}
\caption{{\protect\small (Color online) Transverse momentum distribution of
charged hadrons in the pseudorapidity range }$\left\vert \protect\eta %
\right\vert <1.2${\protect\small \ from minimum bias events in Au+Au (solid
squares) and Cu+Cu (open squares) collisions at }$\protect\sqrt{s}=200$%
{\protect\small \ AGeV for parton scattering cross sections of }$\protect%
\sigma _{p}=3${\protect\small \ mb (left panel) and }$\protect\sigma _{p}=10 
${\protect\small \ mb (right panel). Solid lines are the results\ from Au+Au
collisions that are scaled by }$0.32${\protect\small .}}
\label{dNdPTchg}
\end{figure}

A similar dependence on the size of collision system is also seen in the
transverse momentum distribution of charged hadrons as shown in Fig. \ref%
{dNdPTchg}. For charged hadrons in the pseudorapidity range $\left\vert \eta
\right\vert <1.2$ in minimum bias Au+Au and Cu+Cu collisions at $\sqrt{s}=200
$ \textrm{AGeV} for parton scattering cross sections $\sigma _{p}=3$ and $10$
\textrm{mb}, the transverse momentum spectrum of charged hadrons in Cu+Cu
collisions is again about $0.32$ times that in Au+Au collisions as shown by
the solid lines in Figs. \ref{dNdPTchg}(a) and \ref{dNdPTchg}(b). This
implies that the total transverse energy $\langle m_{T}\rangle $ of hadrons
at midrapidity follows the same dependence on the system size. Since the
initial energy density in heavy ion collisions can be estimated via $%
\varepsilon \sim \langle m_{T}\rangle /(\tau _{0}\pi R^{2})$ \cite{bjorken},
where $\tau _{0}$ is the formation or thermalization time and $R$ is the
radius of the colliding nuclei, the ratio of its values in Cu+Cu and Au+Au
collisions is thus 
\begin{equation}
\frac{\varepsilon _{\text{Cu+Cu}}}{\varepsilon _{\text{Au+Au}}}=\frac{%
\langle m_{T}\rangle _{\text{Cu+Cu}}}{\langle m_{T}\rangle _{\text{Au+Au}}}%
\cdot \frac{R_{\text{Au}}^{2}}{R_{\text{Cu}}^{2}}\sim 0.68.
\end{equation}%
This estimated value is consistent with that from the AMPT model shown in
Fig. \ref{energy}.

\section{Summary}

\label{summary}

Using the \textrm{AMPT} model with string melting, we have studied elliptic
flows in Cu+Cu and Au+Au collisions at RHIC. We find that for both
collisions at same reduced impact parameter and minimum bias collisions, the
elliptic flow of partons in the lighter Cu+Cu collisions is about a factor
of $3$ smaller than that in the heavier Au+Au collisions at same energy,
similar to the ratio of the system sizes. By examining the energy density
and spatial elliptic deformation during the initial stage of a collision, we
further find that their combined effect is similar to the size effect seen
in the elliptic flows in the two colliding systems. A similar ratio is
predicted for the elliptic flows of charged hadrons in minimum bias
collisions of the two systems. Recent preliminary experimental results from
the PHOBOS collaboration \cite{manly} have shown that if the charge hadron
elliptic flow is scaled by the initial eccentricity or spatial elliptic
deformation of participant nucleons, its magnitude becomes qualitatively
similar in both Cu+Cu and Au+Au collisions. How our results are related to
this observation is of great interest and will be addressed in a subsequent
study.

\begin{acknowledgments}
This work was supported in part by the National Natural Science Foundation
of China under Grant Nos. 10105008 and 10575071 (LWC) as well as by the US
National Science Foundation under Grant Nos. PHY-0098805 and PHY-0457265 and
the Welch Foundation under Grant No. A-1358 (CMK).
\end{acknowledgments}

\end{document}